# Electronic Structure of β-Ta Films from X-ray Photoelectron Spectroscopy and First-principles Calculations

Martin Magnuson*, Grzegorz Greczynski, Fredrik Eriksson, Lars Hultman, and Hans Högberg

*Thin Film Physics Division, Department of Physics, Chemistry, and Biology (IFM) Linköping University, SE-581 83 Linköping, Sweden.*

*Corresponding author: martin.magnuson@liu.se

## Abstract

The electronic structure and chemical bonding of β-Ta synthesized as a thin 001-oriented film (space group $P\bar{4}2_1m$) is investigated by *4f* core level and valence band X-ray photoelectron spectroscopy and compared to α-Ta bulk. For the β-phase, the $4f_{7/2}$ peak is located at 21.91 eV and with the $4f_{5/2}$ at 23.81 eV which is 0.16 eV higher compared to the corresponding *4f* peaks of the α-Ta reference. We suggest that this chemical shift originates from electron screening, higher resistivity or strain in the β-Ta film. Furthermore, the *5d-5s* states at the bottom of the valence band are shifted by 0.75 eV towards higher binding energy in β-Ta compared to α-Ta. This is a consequence of the lower number of nearest neighbors with four in β-Ta compared to eight in the α-Ta phase. The difference in the electronic structures, spectral line shapes of the valence band and the energy positions of the Ta *4f*, *5p* core-levels of β-Ta versus α-Ta are discussed in relation to calculated states of β-Ta and α-Ta. In particular, the lower number of states at the Fermi level of β-Ta (0.557 states/eV/atom) versus α-Ta (1.032 states/eV/atom) that according to Mott's law should decrease the conductivity in metals and affect the stability by charge redistribution in the valence band. This is experimentally supported from resistivity measurements of the film yielding a value of ~170 μΩ cm in comparison to α-Ta bulk with a reported value of ~13.1 μΩ cm.

## 1. Introduction

The transition metal tantalum has a property envelope suitable for many applications. As a structural material, the stable α-Ta phase with a body-centered cubic (bcc) crystal structure exhibits a high melting point of 3017 °C that is only exceeded by tungsten, rhenium, and osmium. This property is ideal in filaments for evaporation of, *e.g.*, metals. The high-temperature stability is needed in applications such as substrate holders during thin film growth as well as scanning tunneling microscopy tips. A further advantage of α-Ta is the chemical resistance to corrosion below 150 °C, where the metal is only attacked by hydrofluoric acid and concentrated sulfuric acid [1]. In particular, for thin films, there is an additional phase with a more complicated tetragonal crystal structure called β-Ta that was discovered in 1965 [2]. For sputtered films, β-Ta is the most frequently observed phase, but where growth of α-Ta films [3] as well as mixed α and β-phases are reported [4]. Process development has not yet advanced to the level allowing for deposition of phase pure α-Ta films under industrial conditions.

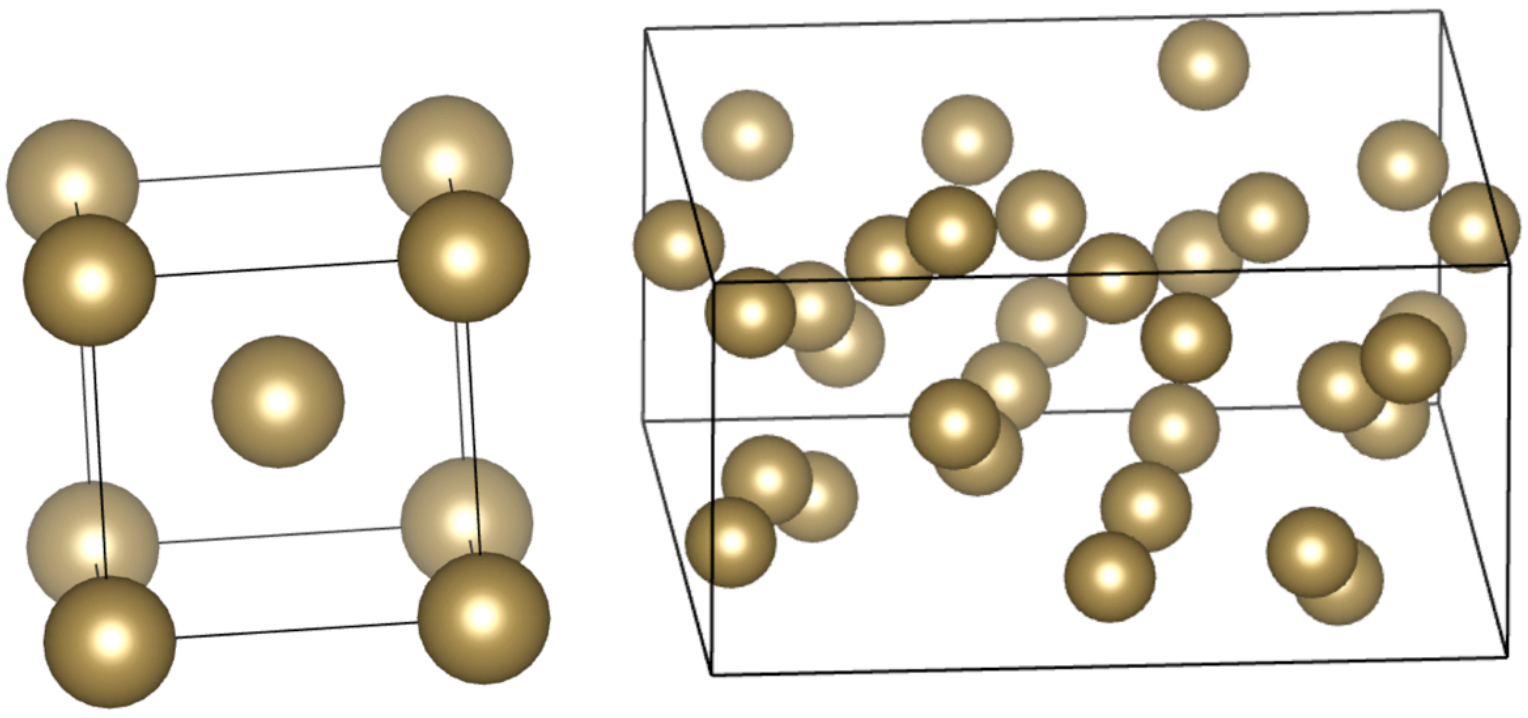

**Figure 1:** Unit cells of a) body-centered close-packed (bcc), α-Ta space group 229 (Im$\bar{3}$m) with atoms in (0,0,0) and (½,½,½) and b) tetragonal β-Ta of space group 113 (P$\bar{4}2_1$m) containing 30 atoms. α-Ta has 8 nearest neighbors with a distance of 2.86 Å and 6 next nearest neighbors at a distance of 3.31 Å, whereas β-Ta has 4 nearest neighbors at a distance of 2.80 Å and 16 next nearest neighbors at a distance of 2.84 Å.

The symmetry of the tetragonal unit cell has been debated as seen from studies by Read and Altman [2], Mills [5], Marples and Shaw [6], Das [7], Catania *et al.* [8], and R. D. Burbank [9]. Predominantly, there are two different space groups suggested for β-Ta; P$4_2$/mnm by Moseley [10] and P$\bar{4}2_1$m as proposed by Arakcheeva [11] [12]. The latter space group is supported by thin film studies by Jiang *et al.* [13] [14] and Högberg *et al.* [15].

Figure 1 shows the crystal structures of α and β-Ta, where the body-centered cubic unit cell of α contains 2 atoms and the β-phase of space group P$\bar{4}2_1$m contains 30 atoms in the unit cell. α-Ta has 8 nearest neighbors with a distance of 2.86 Å and 6 next nearest neighbors with a distance of 3.31 Å [16]. β-Ta has 4 nearest neighbors at a distance of 2.80 Å and 16 next nearest neighbors with a distance of 2.84 Å [17].

Results from thin film growth studies show that the two phases differ with respect to resistivity, α-Ta thin films show a resistivity between 24-50 μΩ·cm [4], while β-Ta has a higher resistivity in the range between 180-220 μΩ·cm [2]. The thin film resistivity of α-Ta is known to be higher than that for bulk-Ta with 13.1 μΩ·cm [18]. In addition, β-Ta is harder compared to α-Ta, which further limits the β phase in applications where metallic properties are requested but promising for applications where higher hardness is needed in tantalum films. Thus, it is necessary to investigate the underlying electronic structure of β-Ta to determine the origin of the property envelope demonstrated by the β-phase and the differences from α-Ta. For β-Ta, studies of the electronic structure by X-ray photoelectron spectroscopy (XPS) measurements including valence band and theoretical density of states (DOS) with density functional theory (DFT) calculations are scarce, and solely for the β-uranium structure [19].

In this work, we apply XPS at the Ta *4f* shallow core levels and the *5d* valence band of a 001-oriented β-Ta film sputtered by high-power impulse magnetron sputtering (HiPIMS) [20]. We investigate the electronic structure to determine the chemical bonding and conductivity properties and compare the results to bulk α-Ta as a reference material. We interpret and support our experimental XPS measurements of the valence band and core levels by first-principle band structure calculations. To test the Mott model, we compare the number of states at the Fermi level with the measured resistivity of the film compared to the α-Ta bulk reference materials. The β-Ta film is particularly useful for characterization of the material's fundamental bonding properties and to identify the structure.

## 2. Experimental details

### 2.1 Thin film growth and characterization

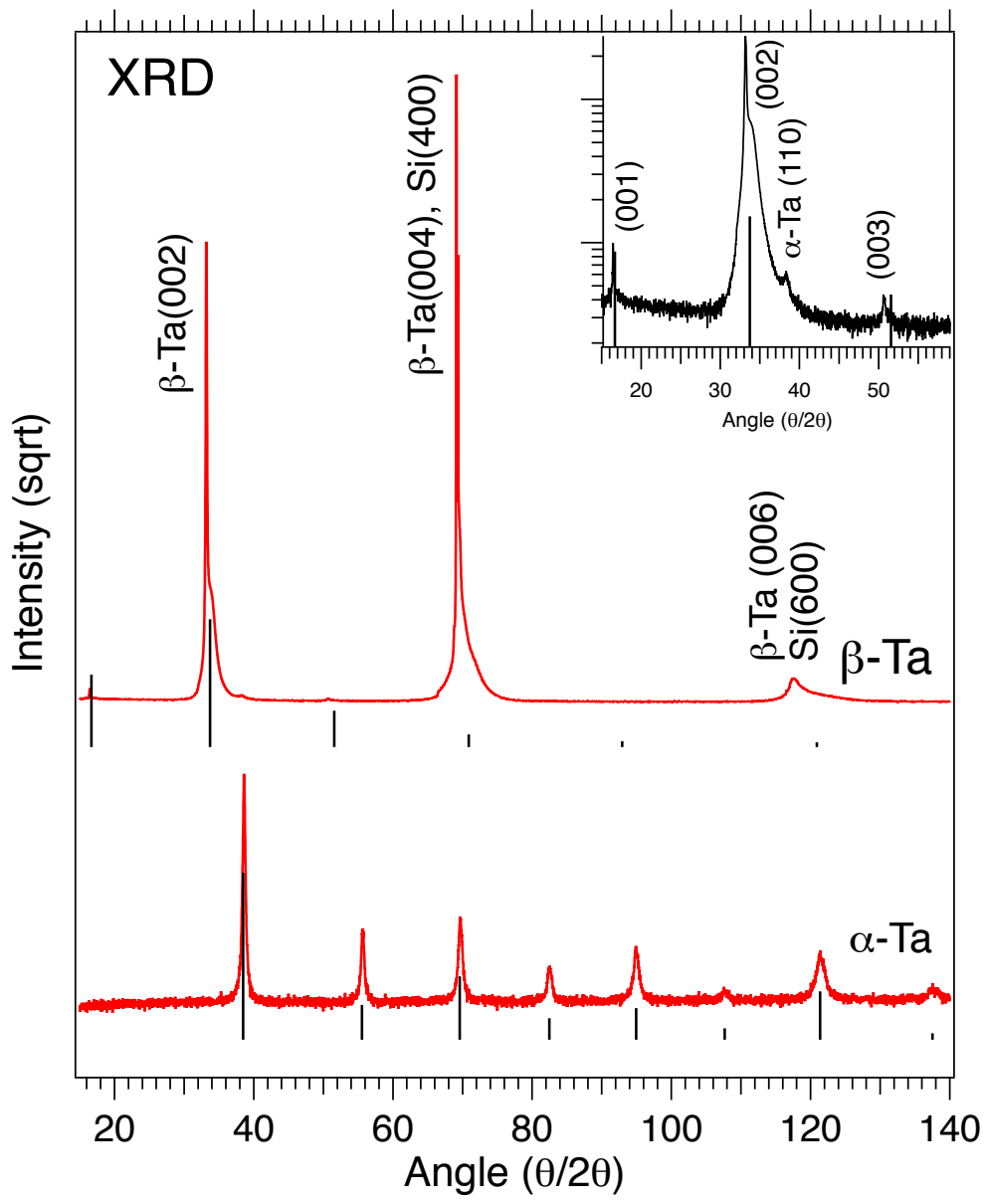

**Figure 2:** XRD patterns from α-Ta bulk (bottom) and β-Ta film (top). The vertical bars represent literature values.

The β-Ta film was sputtered from a tantalum target (99.5% purity) onto Si(100) substrates by HiPIMS in a commercial industrial high vacuum coating system CemeCon AG, Würselen, Germany. The chamber was evacuated to a base pressure below $2.3\times10^{-4}$ Pa prior to deposition. To further lower the impact of the residual gas, sputtering of a titanium target onto a shutter that covered the magnetron was conducted immediately before growth. The β-Ta film was deposited at a fixed target-to-substrate distance of 7 cm with no external substrate heating. The average constant power to the target was set to 3000 W, using a pulse width of 150 µs and a pulse repetition frequency of 300 Hz. The substrate bias was set to -80 V DC and with an Ar partial pressure of 0.42 Pa. The β-Ta film was grown for 570 s, which resulted in a thickness of ~600 nm. For more information about the process conditions and the properties of the β-Ta film the reader is referred to [15]. The investigated α-Ta bulk sample was in the form of a cold-rolled plate.

The chemical bonding structure and the compositions of the films were investigated by X-ray photoelectron spectroscopy (XPS), using an AXIS UltraDLD instrument from Kratos Analytical and analysis with monochromatic Al Kα (1486.6 eV) radiation. The thin film was sputter-cleaned first by 4 keV $Ar^+$ ions for 120 s followed by 0.5 keV $Ar^+$ for 300 s, both at the take-off angle of 20°, to remove adsorbed contaminants following air exposure. The sputter cleaning of the α-Ta bulk sample was more extensive and amounted in total to 2400 s with 4 keV $Ar^+$ and 1200 s with 0.5 keV $Ar^+$ as realized in four cycles each consisting of 600 s (4 keV) and 300 s (0.5 keV). Valence band (VB) *6s* and XPS measurements of the Ta *4d*, *4f,* as well as O *1s* and C *1s* core-level regions were performed at an X-ray incidence angle of 54° and over the area of 300 x 700 µm. The electron energy analyzer detected photoelectrons perpendicular to the sample surface with an acceptance angle of ±15°. The spectra were recorded with a step size of 0.1 eV and a pass energy of 20 eV, which provided an overall energy resolution such that the full-width-at-half-maximum (FWHM) of the Ag $3d_{5/2}$ peak of the sputter-cleaned reference Ag sample was less than 0.5 eV. The peak positions were measured by interpolation by a cubic spline function. In order to avoid confusion associated with the use of the C *1s* peak of adventitious carbon for calibration of the binding energy (BE) scale [21], all core-level spectra were referenced to the Fermi level cut-off which defines zero of the BE axis using a procedure described in ref. [22]. Quantification of the elements in the samples was performed using Casa XPS software (version 2.3.16), based upon peak areas from narrow energy range scans and elemental sensitivity factors supplied by Kratos Analytical Ltd. [23]. The quantification accuracy of XPS is typically around ± 5 %.

The phase contents of the film were characterized by X-ray diffraction (XRD) θ/2θ scans in a Philips powder diffractometer, using Cu Kα radiation (Cu *Kα*, λ=1.54 Å) at 40 kV and 40 mA. XRD pole figure measurements were performed using a Philips X'Pert MRD diffractometer in a parallel beam configuration with a point-focused copper anode source, operating at 45 kV and 40 mA. The primary beam was conditioned using 2x2 mm$^2$ crossed-slits and in the secondary beam path, a 0.27° parallel plate collimator was used together with a flat graphite crystal monochromator. A proportional detector was used for the data acquisition. The detector position was fixed at specific diffraction angles, corresponding to diffraction from β-Ta {002}, {331}, and {413} family of planes, respectively. The pole figure measurements were performed in 5°-steps with azimuthal rotation $0° \leq \phi \leq 360°$ and tilting $0° \leq \psi \leq 85°$ to determine the orientation distribution of the crystals.

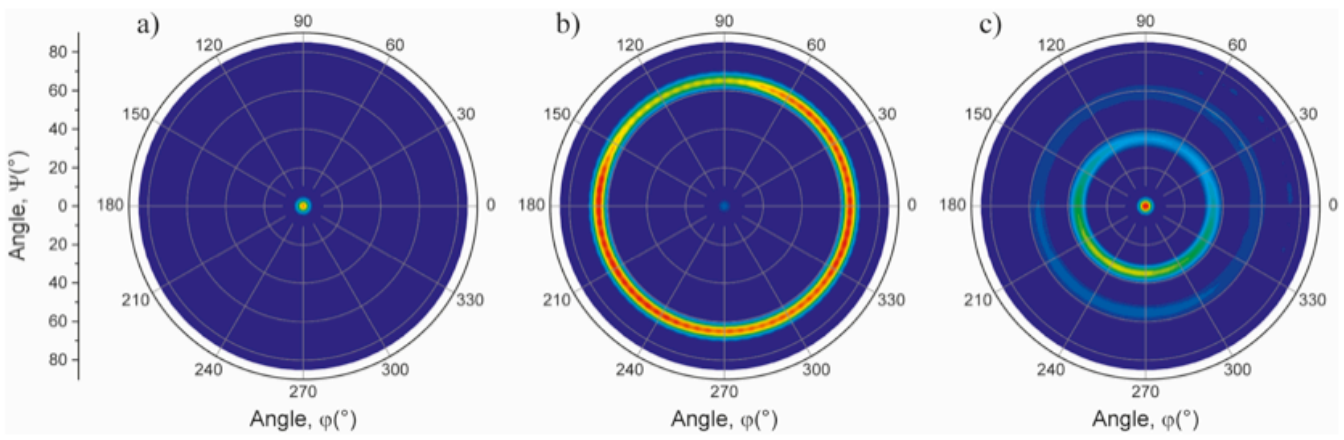

**Figure 3:** XRD pole figure measurements of the a) (002), b) (331), and (413) reflections of the β-Ta film.

### 2.2 Structural model and DFT Calculations

The geometry relaxation was performed using the Perdue-Burke-Ernzerhof (PBE) functional [24] including the Grimme van der Waals density functional theory (DFT)-D2 scheme [25] implemented in the Vienna *ab initio* simulation package (VASP) [26]. For the self-consistent calculations, the WIEN2K code [27] was applied. A 15x15x30 *k*-grid was used for the standard PBE functional for the structure relaxation in VASP (energy cut-off 336 eV) and a 10x10x10 *k*-grid was used for Wien2k functional as the computational cost is significantly higher for the full-potential code.

## 3. Results and Discussion

Prior to measurements of the XPS core level and valence band states, we investigated the structural properties of the α-Ta bulk sample and the β-Ta film, see Figure 2. The XRD pattern at the bottom of Figure 2 was obtained from the bulk sample and demonstrates scattering from all planes characteristic for a bcc metal at the investigated 2θ angles, *i.e.* seen from increasing angles (110), (200), (211), (220), (310), (222), (321), and (400). The 110 peak shows the highest intensity for the ensemble followed by the 211 and 200 as expected for polycrystalline α-Ta. The relatively higher peak intensity of the 110-reflection compared to the other reflections indicate, however, a slight 110 preferred orientation of the α-Ta bulk sample likely due to rolling of the plate. The lattice parameter of α-Ta was determined to 3.3057±0.0002 Å based on a least-squares fit [28], which is in agreement with the literature value of 3.3058 Å [16].

The XRD pattern of the β-Ta film shows three strong diffraction peaks corresponding to diffraction from (002), (004) and (006) lattice planes, showing that the film is 001-oriented. The Si(100) substrate reflections, (004) and (006), are overlapping with the β-Ta peaks. The β-Ta peak positions correspond to a tetragonal structure with space group $P\bar{4}2_1m$ [14] [15]. The identified space group is further supported from the weak, β-Ta 001 and 003 peaks [14] [15], visible using a logarithmic intensity scale, see inset in Figure 2, where also a weak 110 peak from α-Ta is seen at 2θ≈38°.

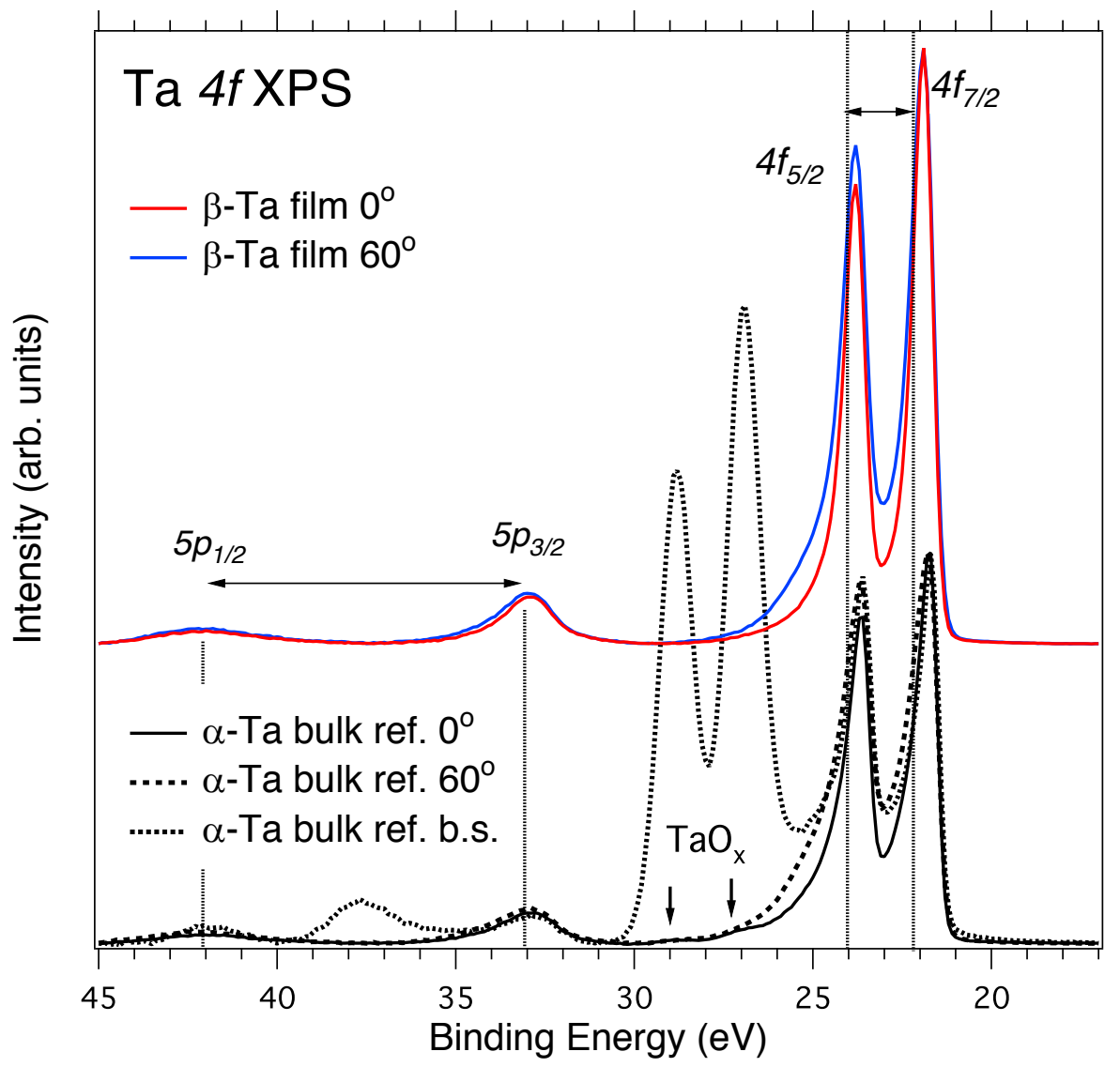

**Figure 4:** Ta *4f* and *5p* core-level XPS spectra of the β-Ta film after sputter cleaning in comparison to α-Ta bulk sample before and after sputter cleaning. The $4f_{7/2,5/2}$ and $5p_{5/2,3/2}$ spin-orbit splittings (1.9 eV and 9.8 eV) are indicated by the horizontal arrows. 0° (in-plane) and 60° (out-of-plane) are the angles from the surface normal to the analyzer when the sample is turned.

To confirm the growth of a 001-oriented β-Ta film, we performed pole figure measurements of the {002}, {331}, and {413} family of planes, see Figure 3 a), b), and c), respectively. For the first diffraction angle the pole figure displays a point of high intensity centered in the middle of the figure where ψ≈0°, which is to be expected for a 001-oriented film. The investigated {331} poles in Fig. 3b supports the 001-orientation from a ring with intensity at ψ≈65° that is close to the theoretical angle of 65.67° between the (002) and (331) planes in β-Ta, and the ring supports a fiber textured growth. The investigated {413} pole in Fig. 3 c) with its ring of high intensity at ψ≈35° finally concludes that the film is 001-oriented as the theoretical angle between the (002) and (413) planes is 35.61°. Our results are consistent with those of sputtered β-Ta-films [4] [29]. From the peak positions of the 002 and 331 reflections, measured by θ/2θ in the growth direction and at a tilt angle of 65.49°, as determined from the 331 pole-figure, the *a* and *c* lattice parameters were determined to 10.339 Å and 5.396 Å, respectively. These values are slightly larger than the literature values for β-Ta with *a*=10.194 Å and *c*=5.313 Å, indicating that the film is under stress.

Figure 4 shows shallow core-level Ta *4f* XPS spectra of the β-Ta thin film sample (top) in comparison to the α-Ta bulk sample (bottom) after Shirley background subtraction and normalization to the $4f_{7/2}$ peak. The angles (θ=0, 60°) are the angles between the surface normal to the analyzer. For β-Ta, the Ta $4f_{7/2}$ and $4f_{5/2}$ peak positions are located at 21.90 eV and 23.81 eV with a 1.91 eV spin-orbit (so) peak splitting. These are higher binding energies than the literature values of 21.6 eV and 23.5 eV, respectively, reported by Fuggle and Mårtensson [30]. The *4f* peak positions of the α-Ta bulk sample are 21.76 eV and 23.66 eV, close to those reported in ref. [30] and the spin-orbit splitting is 1.90 eV. Thus, from our measurements, we observe that for β-Ta, for the $4f_{7/2}$ peak is located at 0.16 eV higher energy compared to the α-Ta bulk sample. In addition, prior to sputtering, the bulk α-Ta bulk sample shows prominent structures at 26.94 eV and 28.84 eV of *e.g.*, tantalum-oxygen bonds ($Ta_xO_y$) [31] following air exposure.

Note that for the α-Ta bulk sample measured at 0° incidence angle, a small contribution of the oxide peaks remains even after sputtering, as indicated by the arrows. The asymmetric tails of the *4f* doublet are less pronounced for the in-plane detection (0°) compared to the out-of-plane detection (60°) when the spectra are normalized to the same background level. This is due to the increased surface sensitivity and the reduced detection area that leads to a reduced signal at 60°. In the 1970's [32] and the 1980's [33], sub-oxides were reported to form broad structures around the Ta *4f* peaks during $Ar^+$ sputtering [34]. Although oxides may contribute to the asymmetry of the tails of the Ta *4f* peaks [35], this effect is relatively small. A contribution from surface core-level shifts (SCLSs) in the spectra would give an estimated shift of 0.3-0.4 eV [36] towards lower $E_b$.

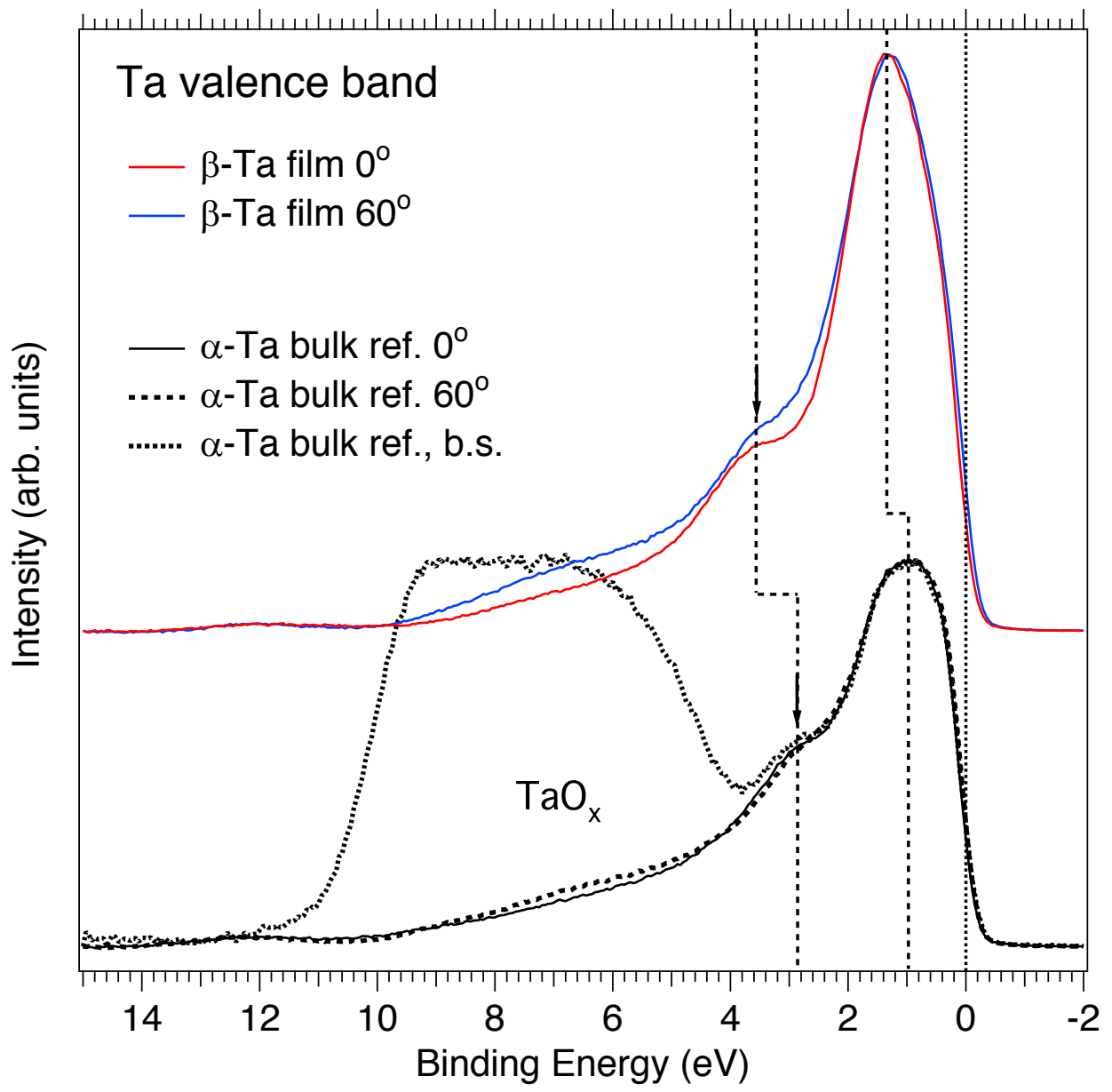

**Figure 5:** Valence-band XPS spectra of the β-Ta film in comparison to bulk α-Ta reference. The shoulder shifts by 0.75 eV to higher energy for the film.

Quantitative XPS analysis of sputtered cleaned samples shows that the β-Ta film has a much lower level of contamination with ~5.8 at% O and ~2.4 at% C compared to the bulk α-Ta bulk sample with ~12.2 at% O and ~14.9 at% C. In the spectrum shown in Fig. 4 of α-Ta prior to sputtering, a high surface oxygen content is confirmed by a broad *5p3/2* Ta-oxide feature located at 37-39 eV. The latter is caused by the significantly higher surface roughness of the bulk reference sample, which prevents efficient surface-cleaning with the $Ar^+$ ions beam. On the contrary, for the β-Ta film, there are no features associated with oxygen after sputtering. The Ta *5p3/2* and *5p1/2* peaks are discernible at 33.00 eV and 42.94 eV, respectively, with a spin-orbit splitting of 9.84 eV. These values are slightly higher than the literature values of 32.7 eV and 42.2 eV, respectively [30]. Note that for the bulk α-Ta sample, there is an additional chemical shift of 0.05 eV towards higher binding energy after sputter cleaning compared to prior to sputtering. This small energy shift is of the same order of magnitude as the instrument energy resolution (±0.05 eV). A higher oxidation state of the Ta atoms due to oxidation would imply a shift towards higher binding energy.

The intensities of the asymmetric tails of the peaks are related to the amount of metallicity in the systems due to the coupling of the core-hole with collective electron oscillations. To evaluate the asymmetric tail toward higher binding energies in each spectrum, a Doniac-Sunjic function [37] corresponding to the electron-hole pair excitations created at the Fermi level to screen the core-hole potential was applied. The Doniac-Sunjic profile consists of a convolution between a Lorentzian with the function $1/E^{\cdot(1-\alpha)}$, where E is the binding energy of each peak and α is a parameter known as the *singularity index* that is related to the electron density of states at the Fermi level [38]. For α-Ta, the singularity index is large (α=0.24) and the *4f7/2*/-*4f5/2* branching ratio is slightly lower (1.14 in-plane and 1.03 out-of-plane) than the statistical value of 1.4 (7/5), signifying high in-plane conductivity. For β-Ta, the asymmetry is smaller than for α-Ta, with a singularity index of only α=0.13 with a branching ratio that is slightly smaller (1.25 in-plane and 1.14 out-of-plane) than the statistical ratio (7/5). The larger singularity index indicates that the number of bands crossing the $E_F$ is higher for α-Ta in comparison to β-Ta, and thus the expected conductivity. For α-Ta before sputtering, the branching ratio is 1.02 indicating lower conductivity than for the sputtered film.

Figure 5 shows high-resolution valence band XPS spectra of β-Ta in comparison to the α-Ta bulk reference material after Shirley background subtraction and normalization to the main peak. As in the case of the *4f* core levels, the detection area is larger for in-plane than out-of-plane. In the energy region up to ~2.50 eV above $E_F$, all the Ta valence band spectra are dominated by Ta *5d* states. The main band of β-Ta is located at 1.35 eV with a shoulder at 3.65 eV below $E_F$ and is associated with bonding states of Ta *6s* character. For α-Ta, the shoulder is located at 2.80 eV, *i.e.*, 0.75 eV lower energy than β-Ta due to the different type of crystal structure. Prior to sputtering, an intense structure located between 4 and 11 eV is due to the O *2p* orbitals of Ta-O bonds in $Ta_xO_y$. Interestingly, the Ta *5d* states have a smaller shift of 0.5 eV compared to the 0.75 eV shift of the *6s* states in the shoulder. In addition, the *5d* band shows a smaller FWHM of 1.9 eV for the β-Ta film compared to the α-Ta bulk material with FWHM=2.2 eV. This is an indication of more localized electrons that leads to higher resistivity in the β-Ta film. Sputtering cleaning of the samples may add to peak broadening. The intensity at the $E_F$ affects the transport properties, *e.g.*, the resistivity of the material.

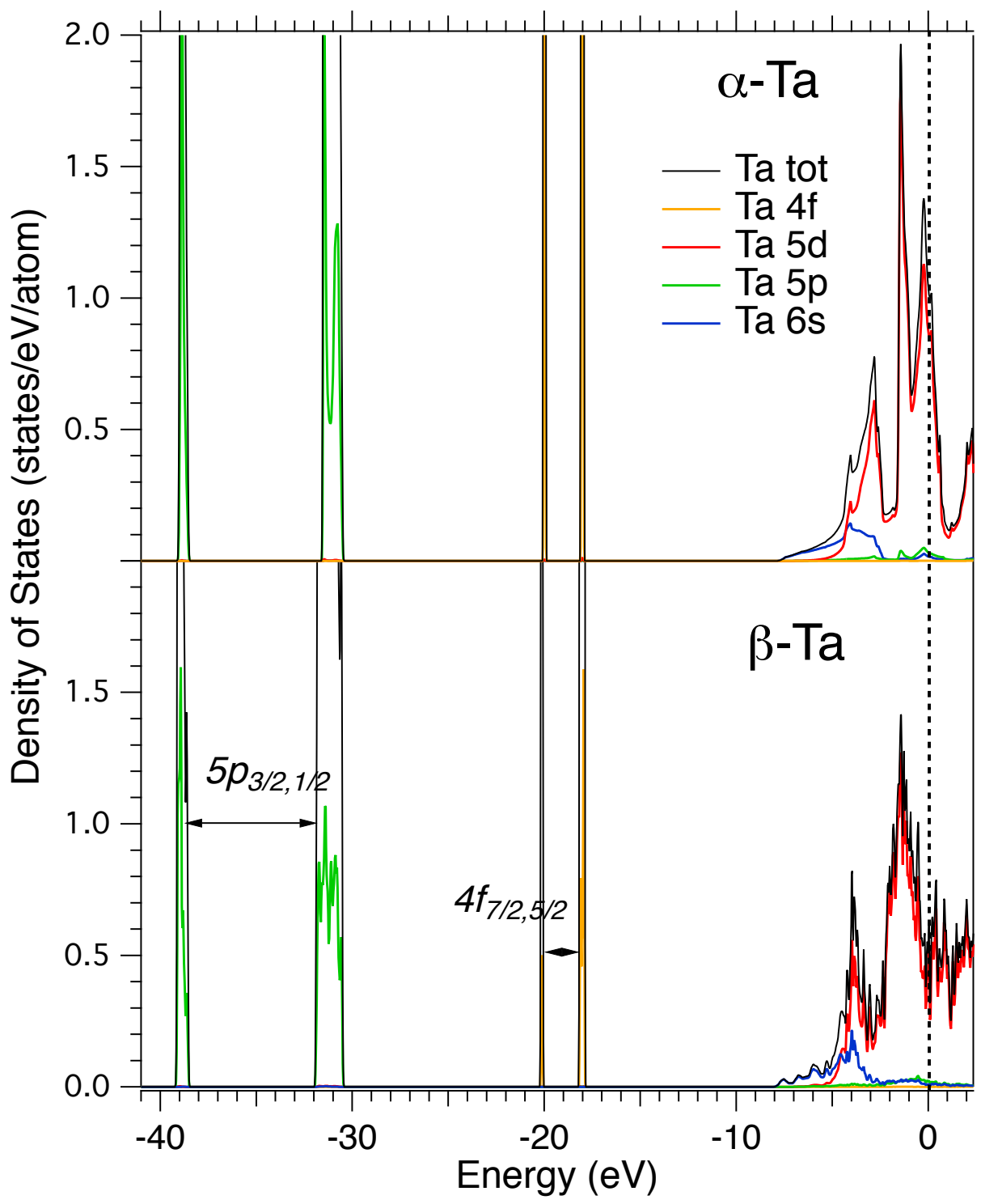

**Figure 6:** Calculated density of states of α and β-Ta. The number of states at $E_F$ is 1.032 and 0.5569 states/eV/atom, respectively.

Figure 6 shows calculated DOS for crystalline β-Ta in comparison to α-Ta. As observed, the states within 5 eV from the $E_F$ are dominated by Ta *5d* states while strong Ta *4f* and *5p* core states are located 18-20 eV (so=2.1769 eV) and 31-39 eV (so=7.5103 eV), respectively. Notably, for β-Ta, the $E_F$ is located in a local minimum that stabilizes the structure and influences the conductivity. On the contrary, for α-Ta, the $E_F$ is located near the top of a narrow *5d* peak of $t_{2g}$ symmetry that should cause more instability of the structure.

**Table I:** Comparison between lattice parameters (LP) and the number of nearest neighbors (NN) and the next nearest neighbors (NNN) for α and β-Ta.

| Sample | Our XRD measurements | literature value | VASP calculations |
|---|---|---|---|
| α-Ta<br>(a-LP) | *a*=3.305 Å<br>NN(8)=2.86 Å<br>NNN(6)=3.31 Å | *a*=3.306Å<br>NN(8)=2.86 Å<br>NNN(6)=3.31 | *a*=3.273 Å<br>NN(8)=2.83 Å<br>NNN(6)=3.27 Å |
| β-Ta<br>(a/c-LP) | 10.339 Å/5.396 Å<br>NN(4)=2.84 Å<br>NNN(16)=2.89 Å | 10.194/5.313 Å<br>NN(4)=2.80 Å<br>NNN(16)=2.84 Å | 10.032/5.309 Å<br>NN(4)=2.72 Å<br>NNN(16)=2.75 Å |

The combined Ta *4f*, *5p* core-levels, and *5d* valence band studies show several interesting effects. A more asymmetric line profile with an associated shift towards lower $E_b$ of the Ta *4f* states in α-Ta in comparison to β-Ta is opposite to the expected high-energy shift of $Ta_xO_y$ surface oxide. An explanation for the chemical shift could be an effect of charge-transfer and redistribution of intensity between the valence band and the shallow core-levels associated with the change in the number of nearest neighbors *i.e.*, the difference in coordination numbers between α-Ta and β-Ta as shown in Fig. 1. We find that the $P\bar{4}2_1m$ space group as proposed by Arakcheeva [11] [12] is in better agreement than the $P4_2/mnm$ space group suggested by Moseley [10].

We attribute the chemical shift to electron screening (less screening leads to higher binding energy) or strain in the β-Ta film. Another reason for the binding energy shift towards higher energy with more symmetric tails for the β-Ta *4f* states compared to the α-Ta states is the higher resistivity of the former phase. The asymmetry of the tails is associated with higher probability of interaction with the electron cloud (screening) when the conductivity is higher. The lattice parameter of the α-Ta bulk target is in good agreement with literature values while both lattice parameters of the β-Ta film are expanded in comparison to the bulk target material by $\Delta a$=+1.42% and $\Delta c$=+1.56%. A potential high-energy shift in binding energy due to an expansion of the lattice is consistent with other XPS measurements [39].

From XRD as shown in Table I, we determined the atomic distances to 2.86 Å and 3.31 Å for the 8 nearest and 6 next nearest neighbors of α-Ta while for β-Ta, the atomic distances are 2.84 Å and 2.89 Å for the 4 nearest and the 16 next nearest neighbors, respectively. These distances are consistent with values calculated from the literature lattice parameters of α-Ta [16], while the distances calculated from the literature lattice parameters [17] are longer for β-Ta, thus indicating tensile stress in the β-Ta film.

The calculated spin-orbit splitting of the Ta *4f* states (2.18 eV) is larger than the experimental value (1.91 eV). The difference between experiment and theory can be ascribed to many-body effects and charge-transfer beyond the one-electron theory [40]. For out-of-plane detection, the asymmetric tails of the Ta *4f* states broaden towards higher $E_b$ due to contribution from core-level shifted surface states [41]. We anticipate that these states can be further resolved using resonant photoelectron spectroscopy at lower excitation energies using synchrotron radiation.

Concerning the valence band, we do not observe any changes due to oxygen contribution. For the β-Ta system, we observe significantly higher intensity for the *5d* valence band further away from the $E_F$ than for α-Ta, indicating lower conductivity for β-Ta. This is also consistent with the larger asymmetry of the Ta *4f* states for α-Ta in comparison to β-Ta. The experimental data show that there is a relatively strong Ta *4f-5d* hybridization and charge transfer towards the valence band. However, this hybridization and charge-transfer is not sufficiently included in the density functional theory.

As shown by the calculations in Fig. 6, the β-Ta phase has fewer states (0.5569 states/eV/atom) at $E_F$ than α-Ta (1.032 states/eV/atom) that affect the conductivity. For comparison, bulk α-Ta has a minimum of resistivity of 13.1 μOhm·cm [18] while the film by Jiang *et al.* [14] is 15 - 60 μOhm·cm [42] [43], and the resistivity for β-Ta is a factor of 3-5 higher 140-180 μOhm·cm [15]. According to Mott's conductivity theory [44], the resistivity is inversely proportional to the number of states at the Fermi level.

## 4. Conclusions

By combining valence band and shallow core-level X-ray photoelectron spectra with electronic structure calculations, we have investigated the highly (001) oriented electronic structure and chemical bonding of β-Ta films in comparison to α-Ta bulk samples. For the investigated β-Ta, the *4f* peak positions at 21.91 eV and 23.81 eV are located 0.16 eV higher binding energies compared to the more asymmetric *4f* peaks of the sputtered and unsputtered α-Ta metal binding energies. The higher binding energies of the β-Ta film in comparison to the α-Ta bulk material are associated with electron screening effects, higher resistivity and strain or potentially the differences in the atomic distances (2.84 Å and 2.86 Å, respectively) of the nearest neighbors. For β-Ta, we find important changes at the Fermi level with a splitting of the *5d* valence band into a pseudogap where the Fermi level is located in a local minimum that stabilizes the structure and minimize the total energy of the system. This is due to a significant redistribution of spectral intensity from the shallow Ta *4f* core levels towards the Ta *5d* valence band states crossing the Fermi level that determines the metallicity.

## Acknowledgements

The research leading to these results has received funding from the Swedish Government Strategic Research Area in Materials Science on Advanced Functional Materials at Linköping University (Faculty Grant SFO-Mat-LiU No. 2009-00971). MM acknowledges financial support from the Swedish Energy Research (no. 43606-1) and the Carl Tryggers Foundation (CTS16:303, CTS14:310). The calculations were performed using supercomputer resources provided by the Swedish National Infrastructure for Computing (SNIC) at the National Supercomputer Centre (NSC) and Center for Parallel Computing (PDC). GG thanks the Knut and Alice Wallenberg Foundation Scholar Grant KAW2016.0358, the VINN Excellence Center Functional Nanoscale Materials (FunMat-2) Grant 2016-05156, the ÅForsk Foundation Grant 16-359, and the Carl Tryggers Foundation grant CTS 17:166.